# Correlation of Structural and Magnetic Properties of $R$FeO$_3$ ($R$=Dy, Lu)

*Banani Biswas[1], Pavel Naumov[1,2], Federico Motti[1,3], Patrick Hautle[1], Marek Bartkowiak[1], Ekaterina V. Pomjakushina[1], Uwe Stuhr[1], Dirk Fuchs[4], Thomas Lippert[1,5], and Christof W. Schneider[1,]\**

[1] Paul Scherrer Institute, 5232 Villigen PSI, Switzerland
[2] Orange Quantum System B.V., Elektronicaweg 2, 2628 XG Delft, The Netherlands
[3] Istituto Officina dei Materiali - CNR-IOM, 34149 Basovizza (Trieste), Italy
[4] Karlsruhe Institute of Technology, Institute for Quantum Materials and Technologies, Kaiserstr. 12, 76131 Karlsruhe, Germany;
[5] Department of Chemistry and Applied Biosciences, Laboratory of Inorganic Chemistry, ETH Zurich, Switzerland

In orthoferrites the rare-earth ($R$) ion has a big impact on structural and magnetic properties in particular the ionic size influences the octahedral tilt and the $R^{3+}$- $Fe^{3+}$ interaction modifies properties like the spin reorientation. Growth induced strain in thin films is another means to modify materials properties since the sign of strain affects the bond length and therefore directly the orbital interaction. Our study focuses on epitaxially grown (010) oriented DyFeO$_3$ and LuFeO$_3$ thin films, thereby investigating the impact of compressive lattice strain on the magnetically active Dy$^{3+}$ and magnetically inactive Lu$^{3+}$ compared to uniaxially strained single crystal DyFeO$_3$. The DyFeO$_3$ films exhibits a shift of more than 20K in spin-reorientation temperatures, maintain the antiferromagnetic $\Gamma_4$ phase of the Fe-lattice below the spin reorientation, and show double step hysteresis loops for both in-plane directions between 5 K and 390 K. This is the signature of an Fe-spin induced ferromagnetic Dy$^{3+}$ lattice above the Néel temperature of the Dy. The observed shift in the film spin reorientation temperatures vs lattice strain is in good agreement with isostatic single crystal neutron diffraction experiments with a rate of 2 K/ kbar bar.

\* Corresponding author: Christof.Schneider@psi.ch.

# 1. INTRODUCTION

The orthorhombic perovskite $R$FeO$_3$ (orthoferrites; $R$: rare earth) is an interesting and technically relevant class of materials for spintronic applications, and has been investigated extensively since the 1960s [1]. Their distinct magnetic properties such as high Neel temperatures ($T_N$ < 600K), a canted antiferromagnetic (AFM) G-type spin structure, a spin reorientation transition (SRT) whose value depends strongly on the ionic size of the rare earth ($R$) element, and an ordering of the rare-earth ions at low temperatures are also very attractive to study their basic magnetic properties [1, 2]. Some of the $R$FeO$_3$ also show ferroelectric (FE) properties [2-4]. Recently, this class of materials gained attention through studies of their low-temperature magnetic field properties [5-7] and the prediction that uniaxial stress will induce FE behaviour even well above room temperature potentially with a large electrical polarization [8, 9]. Such a room temperature multiferroic state would have potential applications in spin based electronic devices.

Strain or externally applied pressure can alter mechanical, chemical, electrical, and magnetic properties of a material [10]. Chemical pressure in which an atom is replaced by another atom is the typical means to introduce strain in a crystalline structure [11-13]. Examples are among the orthomanganites where the change in ionic radius changes the Jahn-Teller distortion thereby changing the antiferromagnetic ground state from an A-type to a spin spiral and finally arriving at an E-type AFM state [14]. This is also reflected in the stability of the crystalline structure where the stable structure will change from orthorhombic to hexagonal with decreasing ionic radius [14]. In thin films epitaxial strain provides control over materials properties by adopting appropriately growth conditions and lattice mismatch between substrate and films to arrive at designed materials properties [15-20]. Chemically induced strain is more closely related to the 3-dimensional (3D) structure of a material, whereas epitaxial strain is more of a 2D type often paired with anisotropic strain characteristics depending on the crystalline symmetry of the substrate and film.

DyFeO$_3$ is one member of the $R$FeO$_3$ family with the space group Pbnm and known to show magnetic ordering-driven multiferroic properties below 4 K [4, 6, 21]. The two magnetic ions form two separate spin lattices and their interaction gives rise to $Fe^{3+}$ – $Fe^{3+}$, $Dy^{3+}$ – $Fe^{3+}$, and $Dy^{3+}$ – $Dy^{3+}$ exchange interactions, which dominate at different temperature regimes. The strong $Fe^{3+}$ – $Fe^{3+}$ interaction results in a G-type AFM ordering along the $a$-axis and the Dzyaloshinskii-Moriya (DM) interaction in a tilting of the oxygen octahedra along the $c$-axis. As a consequence, there is a weak ferromagnetic (wFM) component and A-type AFM ordering below the Neel temperature of the Fe lattice ($T_{N,Fe}$ = 645 K). The interaction between Dy and Fe ($Dy^{3+}$ – $Fe^{3+}$) becomes more relevant with decreasing temperature, which eventually leads to a spin reorientation (SR) in the Fe lattice at around 50K. As a result, the expected Curie-Weiss temperature dependence is changed near $T_{SR}$, and below the G-type AFM ordering switches from the $a$- to the $b$-direction. This goes hand in hand with a change in the magnetic space group from $\Gamma_4$ to $\Gamma_1$ [2]. At around 4K Dy ordering takes place and the spins align antiferromagnetically in the $ab$ plane [6]. Applying a magnetic field below $T_{N,Dy}$ larger than

~2.3T along the $c$-axis [4], the Fe lattice switches back from the $\Gamma_1$ phase to the more stable $\Gamma_4$ phase. Parallel with the change in symmetry a displacement of the Dy atoms along [001] in the unit cell takes place. This magnetic field induced displacement results in a polar unit cell giving rise to FE below $T_{N,Dy}$. As pointed out in [3], applying uniaxial pressure along [110] below $T_{N,Dy}$ reduces the magnetic point group symmetry 222 down to 2 since 2 mirror planes are removed. As a consequence, a FE polarization is introduced below $T_{N,Dy}$ with the FM moment parallel [001] $T_{N,Dy}$ [3]. There is also a report where a switchable electric polarization has been observed along [100] above $T_{SR}$ when poling was carried out across the SRT while cooling and measured while warming. It is concluded that a magnetic ordering of the $R$ ions is not essential for inducing FE but the combination of a paramagnetic $R$ ion and the $\Gamma_4$-phase (wFM) is sufficient [22, 23].

In this paper, we study the effect of growth induced strain on the magnetic properties of the nominally multiferroic material DyFeO$_3$ and LuFeO$_3$ as thin films by growing them epitaxially strained on (010)-oriented YAlO$_3$ substrates with a compressive in-plane strain. For the magnetic characterization, the temperature dependence of the magnetic moment ($M$) was measured for all three main crystallographic axes as well as the moment vs magnetic field up to 7 T between 1.8 K and 390 K. To compare the effect of compressive strain and pressure on the magnetic thin film properties neutron diffraction measurements were performed to probe the temperature dependence of the SR of a DyFeO$_3$ single crystals in a He gas pressure cell with an isostatic pressure up to 5kbar (0.5 GPa). Due to the experimental set-up, the single crystal data are equivalent to thin film experiments.

## 2. Experimental methods

Epitaxial thin films of DyFeO$_3$ ($a$=5.302 Å, $b$=5.598 Å and $c$=7.623 Å) and LuFeO$_3$ ($a$=5.2176 Å, $b$=5.5556 Å and $c$=7.5749 Å) are grown on (010) oriented YAlO$_3$ single crystalline substrates ($a$=5.18 Å, $b$=5.330 Å, and $c$=7.375 Å) by pulsed laser deposition using a KrF excimer laser ($\lambda$ = 248nm, 3 Hz) with an in-plane lattice mismatch between (010) DyFeO$_3$ and a (010) YAlO$_3$ substrate of -2.36% along the $a$- and -3.36 % along the $c$-axis, -5.03 % along the $b$-axis [24]. The lattice mismatch for LuFeO$_3$ is -0.73 % along the $a$-, -4.23% along $b$-, and -2.71 % along the $c$-axis. YAlO$_3$ is a non-magnetic and non-ferroelectric insulating substrate and therefore suitable to study magnetic and dielectric/FE properties of multiferroic films on the same substrate. The laser beam is focused onto a sintered ceramic target with a spot size of 1.4 x 1.4 mm$^2$ with the laser fluence adjusted to 2.0 J cm$^{-2}$. The substrate is located on-axis to the plasma plume with a distance of 5 cm from the target. Deposition was performed in an O$_2$ background at 0.33 mbar with the substrate heated to 700°C by a lamp heater [24]. Five films of different thickness have been deposited subsequently (13, 27, 51, 78, 115 nm) to study the thickness dependence of crystalline and magnetic properties. Other films discussed were prepared at different preparation runs with the aforementioned deposition conditions.

The structural quality of the powder and single crystal, and the thickness of the films after the deposition has been monitored by X-ray diffraction (XRD) and -reflection (XRR) respectively using a Seifert 3003 PTS and a Bruker D8 Advance four-circle x-ray diffractometer with a Cu $K_{a1}$ monochromatic x-ray source. Θ-2Θ scans have been used to monitor the crystalline quality of all samples, reciprocal space maps to determine in-plane lattice constants and to evaluate qualitatively strain in these films, XRR to determine the thickness of the films after the deposition.

The temperature dependent magnetization was investigated by a commercially available SQUID magnetometer (Quantum Design, MPMS® 3). It enables temperature sweeps between 1.8 - 400 K with a magnetic field of up to 7 T parallel to the measurement direction. The evolution of the magnetic moment (*M*) with temperature has been analysed by the following measurement protocol. First, the sample is cooled down without applied field to 3 K. The zero-field cooling (ZFC) *M*(*T*) measurement is taken during heating between 3K and 300 K in the presence of a small field applied along the measurement direction. After each measurement cycle, the film was kept at 390 K for 5 minutes to erase as much as possible trapped fields. Normally, the annealing temperature should be higher than $T_{N,bulk}$ ~ 645 K. Due to instrumental limitations, the maximum temperature which can be reached inside the SQUID is ~400 K. Although, this temperature is not enough to remove the trapped stray fields inside the film completely, increasing the temperature up to ~400 K will certainly increase thermal fluctuations of the magnetic spin and minimize trapped field since $T_N$ for strained films is expected to be considerably lower than $T_{N,bulk}$ [6]. After this annealing process, the samples are cooled down with an applied field and measurements are taken during cooling (field-cooled cooling, FCC) and heating (field-cooled heating, FCH).

Neutron diffraction studies were performed at the spallation neutron source SINQ, Paul Scherrer Institute (Villigen, Switzerland). Single crystal neutron diffraction experiments were conducted at the thermal triple-axis spectrometer EIGER with $k_f$ = 2.66 Å$^{-1}$ (λ = 2.36 Å) [25] using a pyrolytic graphite filter to eliminate higher-order neutrons. The zero-field experiments were done in a He-cryostat to reach temperatures down to 1.5 K with the sample mounted in the (0*kl*) scattering plane. For the pressure dependent measurements, a helium gas pressure cell [26-28] has been used with a max reachable pressure of 5 kbar to study the spin reorientation transition at 40K in a single crystal under uniaxial pressure conditions between 1.5 K and 100 K. In the temperature regime of interest, the gas-pressure cell has the advantage that the pressure medium He is still liquid and the applied pressure therefore truly isostatic. This enables to understand orientation dependent measurements correctly. At temperatures below 20K, He solidifies at the selected pressures and the required isostatic pressure conditions would not be fully met. The cell design is a monobloc thick-wall cylinder cell with a Ø 16 mm bore similar to a design described by Aso *et al.* [26]. To minimize neutron absorption, the aluminium alloy (Al 7049A-T6) is used to build the cells containing an active volume of ~0.010 *l*. For the measurements, the cell is mounted on a dedicated cryostat insert inside a cryogenic system suitable to be used at the EIGER beam line. For each pressure value, a *k*-scan was done at 80K and 5K to determine the background contribution for the appropriate background correction and lattice constants.

## 3. EXPERIMENTAL RESULTS

### 3.1 Crystallographic structure of (010)-oriented DyFeO$_3$ and LuFeO$_3$ thin films

The crystalline quality of the DyFeO$_3$ films with five different thicknesses was studied using XRD. All films are (010)-oriented and show a good crystalline quality with the FWHM for the ω-scans of ~0.02°. The strain for the (010) out-of-plane direction is tensile for all film thicknesses measured (Fig. 1a) and thickness dependent. For thinner films of around 10 nm, it is of the order 2-3% and decreases down to < 1% for films with a thickness of 100nm or more. Films with a thickness between 15 – 20nm grow coherent as shown exemplary in Fig. 1b for a 13 nm thin film with the in-plane axes aligned macroscopically with the substrate (Fig. 1b). Thicker films show a relaxation towards the bulk DyFeO$_3$ Bragg position, but some remnants of the initial coherent layer can still be found effectively forming a two-layered film system. Evaluating the DyFeO$_3$ film lattice parameters we can distinguish three regimes ((i) highly strained, (ii) strained, (iii) largely relaxed) with a change-over in the 40nm range film thickness. Two film thickness for LuFeO$_3$ have been studied. A fully coherently grown sample with 21 uc (11.7nm) and one sample with 21nm. Like for DyFeO$_3$, the strain in the *b*-direction is tensile and smaller than for DyFeO$_3$ and the in-plane [001] is compressive. For the *a*-direction, the 11.7nm film has a compressive strain, whereas the 21nm film is largely relaxed in this direction and lightly tensile. From the reciprocal space maps, both films look coherently grown with strain fields developing for the thicker film (Fig. 1a). This also holds for the (150) Bragg-peak. Interesting to note is the fact that the compressive strain along [100] for the 21uc film is much larger (-2.2%) than one would expect for a nominal -0.73% substrate/bulk lattice mismatch followed by a 0.2% tensile strain for the thicker film. The reason for this finding is not fully understood.

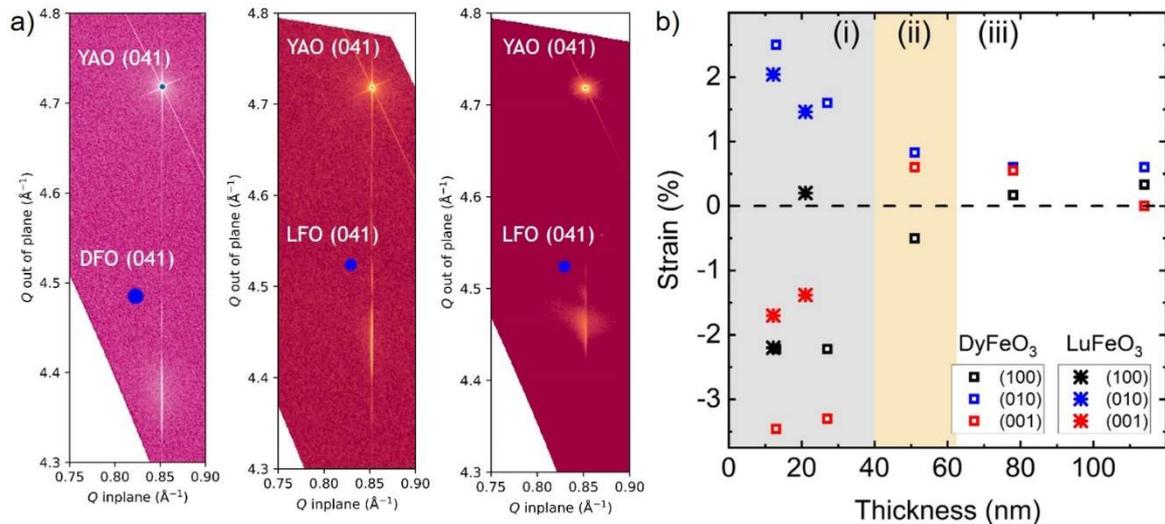

**Figure 1:** a) Reciprocal space maps of the (041) reflections of a 13 nm (010) DyFeO$_3$ film, a 11.6nm and 21nm (010) LuFeO$_3$ film grown on (010)-YAlO$_3$. The blue dot in each map indicates the expected position of bulk DyFeO$_3$ or LuFeO$_3$. b) Crystalline axes dependent strain vs. film thickness for (010) DyFeO$_3$ and LuFeO$_3$ grown on (010) YAlO$_3$.

For a thickness thinner than 40nm, films are highly strained with a constant unit cell volume of ~219 Å$^3$, whereas films thicker than 40nm are partially strained and the constant unit cell volume (~229 Å$^3$) is similar to unstrained DyFeO$_3$ (~226 Å$^3$). This is also reflected when calculating the out-of-plane strain, which is for the 13nm and 27nm film ~2.6% and ~1.6% and for the thicker films ~0.6%. The experimental situation for the two LuFeO$_3$ films with respect to the unit cell volume is a different. The volume of unstrained LuFeO$_3$ (~219.6 Å$^3$) is smaller than the volume of the fully coherently grown LuFeO$_3$ film (~223.8 Å$^3$), indicating a Poisson ratio < 0.5.

## 3.2 Magnetic Properties

The Fe lattice in bulk DyFeO$_3$ orders antiferromagnetically with the spin pointing along the *a*-axis. At ~50 K [4-6, 29, 30] the Fe spins reorient themselves from the *a*- to the *b*-direction as a consequence of the increase in the Fe$^{3+}$-Dy$^{3+}$ interaction and the space group changes from a $\Gamma_4$ to $\Gamma_1$. With the change in magnetic symmetry, the wFM component of $\Gamma_4$ vanishes below $T_{SR}$. The change and direction of the Fe spin under isostatic pressure can be verified by neutron diffraction by measuring the temperature dependence of the (031) Bragg peak of a DyFeO$_3$ single crystal.

### 3.2.1 Temperature and isostatic pressure dependent single crystal neutron diffraction

The magnetic properties of DyFeO$_3$ are known to be very sensitive to external stimuli due to the very small interaction energies involved [7]. This seems to be a general characteristic for the *R*FeO$_3$ family. So far, mainly magnetic field dependent measurements have been done revealing strong changes in the magnetization, in particular at low magnetic fields at and below $T_{SR}$. Looking into the pressure dependence of the spin reorientation is therefore of interest since an applied pressure will change the bond-length and therefore the interaction strength. Tensile strain, i.e., quasi negative pressure is expected to weaken the interaction whereas compressive strain (positive pressure) should strengthen it. In the latter case, $T_{SR}$ should increase. Isotropic pressure dependent single crystal neutron diffraction measurements can hence be used as a frame of reference for magnetic thin film data in particular for the compressively in-plane strained thin films. The aim of these experiments is therefore to distinguish between pressure-induced uniaxial effects on the magnetic ions e.g., a change in $T_{SR}$, and anisotropic, lattice related effects on magnetic signatures as they potentially occur in strained thin films.

The temperature dependence of the (100), (010) and (001) lattice constants at ambient pressure conditions are shown in Fig. 2(a) as determined using the Morpheus beam line at SINQ, PSI [7]. They vary remarkable little between room temperature and 2K. The difference for the *a*-direction is 0.0023 Å, for the *b*-direction 0.0074 Å and the *c*-direction 0.011 Å. Applying a pressure up to 3.9kbar, we measure at 8K and 100K a pressure induced change of the lattice parameter of 0.01Å/kbar along [100] and 0.015 Å/kbar along [001], the [010] direction has not been measured. The pressure induced changes for the lattice constants are larger than the temperature dependent changes of the lattice constants shown in Fig. 2(a). They are also larger than the changes in lattice parameter obtained from pressure dependent Raman measurements [31].

To characterize the temperature dependence of the SR, the maximum peak intensity of the magnetic (031) Bragg peak was measured between 5 K and 100 K at ambient pressure [7] and for a pressure window between 200 bar (the smallest reachable pressure in the cell) and 3.9 kbar (Fig. 2(b, c)). For this scattering geometry, the spin propagation vector $\mathbf{q}_{Fe}$ is along [001] above $T_{SR}$ and along [010] below. In Fig. 2(b), an ambient pressure measurement is compared to the 200 bar measurement. Both measurements show a constant count rate above $T_{SR}$, indicating that all measured moments are aligned along [100] [7] and the sharp drop at $T_{SR}$ is unchanged with increasing pressure. Below $T_{SR}$ the additional step in intensity between 10 and 20 K is still measurable like for the ambient pressure measurement, albeit, the intensity under pressure is lower. This indicates already even a small pressure, small compared to the nominal pressure associated with a film-substrate lattice mismatch, is having a measurable effect on the spin alignment in an orthoferrite system. This observation is also in line with the sensitivity of $DyFeO_3$ with respect to pressure as reported in [3]. With increasing pressure, the sharpness of the SR-transition remains unchanged and the intensity below $T_{SR}$ reaches a value associated with the background of the (031) peak at 3.9 kbar and 6 K (Fig. 2(c)). These measurements show, that the Fe-spins flip almost completely from [100] to [010] at $T_{SR}$ and the increase of $T_{SR}$ with pressure takes place at a rate of 2 K/ kbar (20K/GPa). This rate is similar to measurements performed on hematite where a rate of 3 K/ kbar on $T_{SR}$ in a hydrostatic environment up to 0.6 GPa was reported [32, 33]. However, measurements in a non-hydrostatic environment up to 10 GPa showed a change in $T_{SR}$ between 0 and 7GPa of ~12K and deviatoric stress was mentioned as a likely reason for this very different pressure dependent behaviour. Like for an ambient pressure measurement, there is no sign of a Dy-ordering along [001] at low temperatures. We therefore conclude to have a Dy-ordering below 5 K in the $ab$-plane in the presence of an isostatic pressure [7].

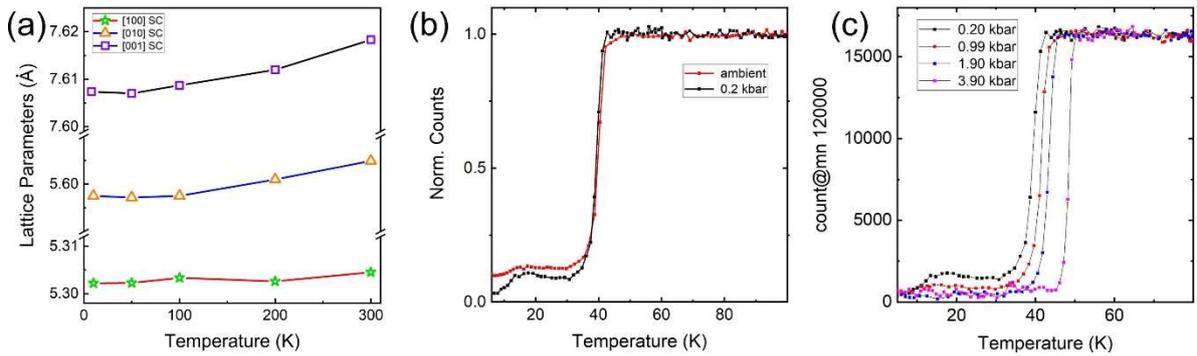

*Figure 2:* a) Temperature dependence of the $DyFeO_3$ single crystal lattice constants at ambient pressure b) Magnetic (031) diffraction peak of a $DyFeO_3$ single crystal measured at ambient pressure and 0.2kbar. c) Pressure dependence of the magnetic (031) diffraction peak of a $DyFeO_3$ single crystal. The measured intensity has been recorded with a monitor of 120000.

### 3.2.2 Temperature dependence of the magnetic moment in zero field for thin films

We first present zero field cooled $M(T)$ measurements for the 115 nm thick film between 300 K and 2 K for the $c$-direction (Fig. 3 a). The thickest film has the most relaxed structure compared with thinner films and $M(T)$ is expected to be similar to single crystal data. For a single crystal two magnetic transitions are observed, one at ~6 K and one at ~41 K, with the first one being $T_{N,Dy}$, the latter $T_{SR}$ where the Fe-spins flip from the $b$- into the $a$-direction as determined from neutron diffraction measurements [7]. Characteristic for this SRT is a width spanning only a few K within most Fe-spins change direction.

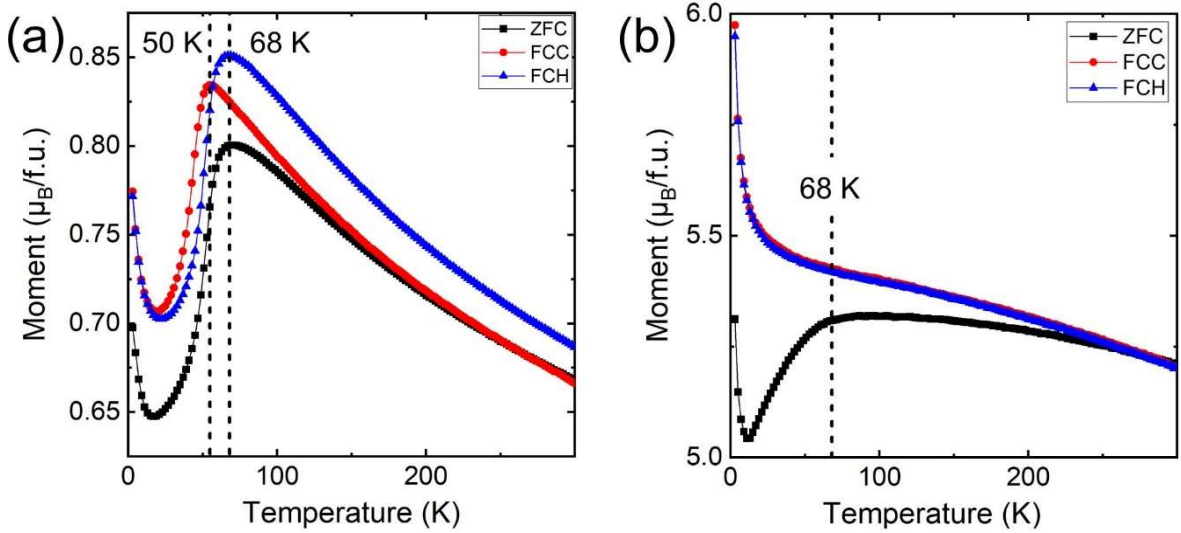

*Figure 3:* a) ZFC and FC measurements of a 115 nm, (010) oriented DyFeO$_3$ thin film along [001] with a magnetic field of $H = 0.1$ T. The dashed lines show the transition temperatures at 68 K and 50 K. b) ZFC and FC measurements of a 13 nm, (010) oriented DyFeO$_3$ thin film along [001] with a magnetic field of $H = 0.1$ T. The dashed line shows the transition temperature at 68 K.

The ZFC measurement for the 115 nm film along [001] shows an Fe spin reorientation transition at ~68K and the beginning of a Dy ordering at ~17K (Fig. 3a). Whereas Dy ordering is observed for the $a$- and $b$-direction at ~4.5K, there is no signature of $T_{SR}$ along the $a$- and $b$-direction at any temperature above. In analogy to a single crystal [7] there is AFM Dy-ordering in the $ab$-plane below $T_{N,Dy}$, however, the low-temperature $M(T)$ upturn along [001] is unlike known properties for a single crystal with a potential ordering temperature well below 1.6 K. For a single crystal the measured moment along [001] at $T_{SR}$ is ~0.15 µ$_B$/f.u., for the film it is ~0.8 µ$_B$/f.u. and therefore considerably larger. At $T_{SR}$, the change in slope along the $c$-axis is rather broad and occurs over a large temperature window in stark contrast to the abrupt drop in moment for the single crystal. Assuming the direct analogy to the single crystal, the Fe-spin will point along [100] above $T_{SR}$ and along [010] below.

$M(T)$ measurements for the highly strained 13nm thick film shows along [001] a change in curvature at ~68 K and the signal drops very slowly with decreasing temperature until at ~13 K the moment rises again (Fig. 3b). Like for the 115 nm film, the low temperature rise in moment is taken as evidence of a starting Dy ordering with an AFM Dy ordering below 4.5 K in the $ab$-plane and the measured moment at $T_{SR}$ reaches ~5.28 µ$_B$/f.u. along [001]. ZFC $M(T)$ measurements for films with a thickness of 27, 51 and 78nm showed similar properties as shown for the thinnest and thickest films with a Dy-ordering between 4 and 5K for [100] and [010]. The measured moments at SRT increase with decreasing thickness and values are between the thickest and thinnest film.

### 3.2.3 Field cooled thin film measurements

Trapped fields in single crystals can have a pronounced influence on $M(T)$ resulting in additional steps and humps at temperatures typically above $T_{SR}$ [7]. The origin of the field sensitivity are the small interaction energies, in particular along the soft magnetic axis and $T_{SR}$ is systematically shifted to lower temperatures with increasing field. In the case of relatively thick films (>40nm), the onset of the transition associated with the SR for FC measurements can vary significantly. An example is shown in Fig. 3a where FCC and FCH for the 115nm thick film is shown. For the FCC measurement, $T_{SR}$ is shifted to lower temperatures by 18 K and the value of $M$ has increased. The same finding applies for the FCH measurement where the values for $M$ are larger compared to the ZFC measurement and $T_{SR}$ is back to 70 K. Likewise, the opposite can happen. Trapped fields shift $T_{SR}$ up, in some cases even well beyond 100K. Larger values for $M(T)$ for FC measurements is a general observation for films of all thicknesses. For films with a thickness of less than 40 nm, the SRT measured along the $c$-direction is not measurable anymore (Fig. 3b).

The ZFC and FC (heating or cooling) curves along [001] show a divergence at low temperature and they converge near room temperature (~250 K). Besides, the net moment during any field cooled measurement shows a higher value at a fixed temperature than the moment measured during ZFC for the temperature range below 250 K. The evolution of the interaction between the Dy and Fe lattice with applied magnetic field is a possible reason for this behaviour. The magnetic interaction between $Dy^{3+}$ and $Fe^{3+}$ has been reported to induce ordering of the Dy lattice for SC $DyFeO_3$ at ~210 K along the $c$-axis and the direction of the $Dy^{3+}$ spins depends on the direction of the applied magnetic field [5].

Near room temperature, the $Dy^{3+}$ spin system is the paramagnet and thermal energy prevents the ordering of the Dy lattice [5]. As the thermal fluctuations are reduced with decreasing temperature, the induced magnetic ordering of the $Dy^{3+}$ spin lattice increases due to a dominant an increase in $Dy^{3+}$-$Dy^{3+}$ interaction. During the ZFC measurements, the sample is cooled down to 3K in the absence of a field. The spin state of the Fe lattice at 3 K is in the $\Gamma_1$ phase. The induced weak magnetic ordering of the Dy lattice is linked with the Fe lattice in the $\Gamma_4$ phase. Therefore, there is no or a small contribution from the $Dy^{3+}$ ion to the total moment below $T_{SR}$. For FC measurements, the absence of a SR would indicate that the Fe lattice stays in the $\Gamma_4$ phase below $T_{SR}$ and the $Dy^{3+}$ spin-alignment is linked to $\Gamma_4$. Therefore, the combined contribution of the $Dy^{3+}$ and $Fe^{3+}$ moments increase the total moment for these FC measurements, which leads to the separation between the ZFC and FC curves.

### 3.2.4 $M(H,T)$ and double step hysteresis loops

Isothermal field-dependent magnetic measurements, $M(H)$, along all three major axes have been performed on these films to study the spin interaction of $Fe^{3+}$ - $Fe^{3+}$, $Dy^{3+}$ - $Fe^{3+}$ or $Dy^{3+}$ - $Dy^{3+}$ between 1.8 K and 390 K. For single crystal DyFeO$_3$, $M(H)$-loops along [001] show a small magnetic hysteresis at low fields and a saturation moment of 0.15 µ$_B$ at 50 K with the hysteresis closing at $T_{SR}$. No hysteresis is observed for the other directions [7]. The closing hysteresis at $T_{SR}$ is the signature of the change in spin direction and the change in magnetic symmetry from $\Gamma_4$ to $\Gamma_1$. In contrast to a single crystal, a FM hysteresis for the 115nm film is observed for all three crystalline directions at room temperature (Fig 4a) and also measured for all films of different thickness between 1.8K and 390K. Exemplary, $M(H,T)$ along [100] for the 115 nm film is shown in Fig. 4b. It is clear, the hysteresis loops along [001] are sustained even well below $T_{SR}$ with an increase in remanence and coercivity with decreasing temperature. Since the change from $\Gamma_4$ to $\Gamma_1$ compensates the FM contribution the presence of a FM hysteresis loop between 1.8K and 390K indicates that the $\Gamma_4$ phase exist above and below $T_{SR}$ irrespective of the films thickness.

For $M(H)$-loops measured along the in-plane [100] and [001] directions, double step hysteresis (DS) loops are observed as shown in Fig. 4a, and for films of different thickness (Fig. 4c) exemplary shown at 80 K along [001]. Similar to the FM hysteresis-loops, the DS loops are observed for all film thicknesses and within a temperature range between ~10 K and 390 K (Fig. 4b). Common is also the large magnetic field (up to ~1 T) where the step takes place. The DS-loops were observed for the in-plane [100] and [001] directions and not for [010]. Below ~10 K, Dy ordering starts to take place and the DS becomes a single hysteresis loop indicating a substantial increase in the $Dy^{3+}$-$Dy^{3+}$ interaction. The origin of a DS in a potential two-spin system is the coupling between two ferromagnetic layers/lattices with different coercive fields [34-38]. Measuring $M(H,T)$ for LuFeO$_3$ films with $m_{Lu}$=0 and similar strain properties like DyFeO$_3$, we find for the *a*-direction of the coherently grown, 21 uc thick film a DS hysteresis and a single loop for [001] between 50 K and 390 K albeit with a smaller switching field at DS compared to DyFeO$_3$ films. The observation of a DS hysteresis for this single spin system cannot originate from a coupling of spin systems but must instead arise from a change of the canting angles and a distortion of the Fe-O octahedra as a consequence of growth induced compressive strain. This is equivalent to a biaxial magnetic anisotropic strain in a unit cell [39-42], here -2.2% and -1.7% for [100] and [001], respectively, giving rise to a stronger Fe-O octahedra tilting and hence a canting along both in-plane directions. Since strain values for LuFeO$_3$ film are similar to DyFeO$_3$, strain induced DS in DyFeO$_3$ hysteresis loops can be expected. In DyFeO$_3$ films, however, the DS-loops occur largely as result of a coupling between the ferromagnetically ordered Fe lattice and the induced ferromagnetically ordered Dy lattice [5] since coercive field values are considerably larger as compared to LuFeO$_3$ because both moments add up. Further, for DyFeO$_3$ films the DS is present for both in-plane directions whereas for LuFeO$_3$ films the DS is measured only along one crystallographic direction. The magnetic field $H_{C2}$ at which the step occurs increases monotonically from 0.04T at 390K to 0.11T at 50K (Fig. 4d; insert). A schematic representation of the spin alignment of the Fe and Dy at different field values is illustrated in Fig 4e with the arrows indicating the direction of the spin alignment for the Fe and Dy lattices. However, the spin system in our films switches from a FM-FM coupling to a FM-AFM coupling below $T_{N,Dy}$ as Dy orders antiferromagnetically in the *ab* plane. As a result, with the onset of Dy ordering the DS vanishes and the centre of the hysteresis is shifted leading to an exchange bias effect.

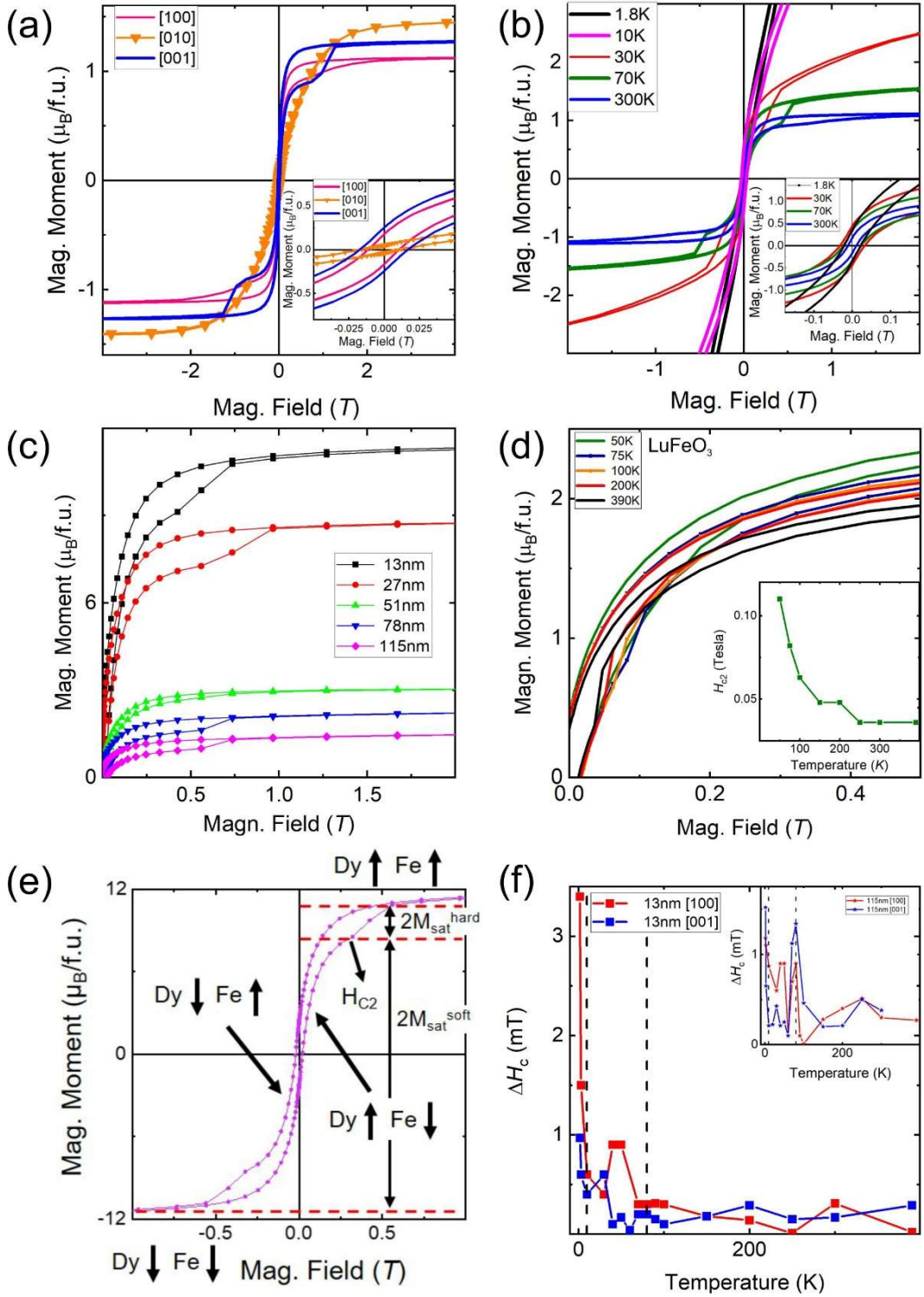

*Figure 4:* a) $M(H)$ at RT along [100], [010] and [001] for the 115nm film; insert FM hysteresis. b) $M(H)$ between 300 and 1.8K of 115 nm film along [100]. c) Double step hysteresis along [001] @ 80K for all film thicknesses. d) $M(H,T)$ along [100] for a 21uc thin (010)-oriented $LuFeO_3$ thin film; insert: Coercive field ($H_{C2}$) vs temperature. e) Explanation of the double step hysteresis @50K. The arrows show the direction of the spin alignment for the Fe and Dy magnetic lattices. $M_{sat}^{hard}$ and $M_{sat}^{soft}$ represent the saturation moment of the hard magnetic ordering (Fe) and soft magnetic ordering (Dy) respectively. $H_{C2}$ represents the coercive field associated to the hard magnetic ordering. f): Exchange bias field $\Delta H_c(T)$ along [100] and [001] for the 13nm film and 115nm film (insert).

The applied magnetic field can change the spin alignment in the Dy spin lattice above $T_{SR}$ from an AFM to a FM configuration. It is therefore expected to measure an exchange bias effect, in particular at low temperatures since a $DyFeO_3$ films seems to be a natural FM-AFM spin multilayer where a pinning of the FM with the AFM spin lattice can occur. In Fig. 4f the temperature dependence of the exchange bias field $\Delta H_c$ for [100] and [001] for the 13nm and 115nm thick films is shown. Above 150K, for both film thicknesses there is no clear exchange bias effect to be measured. This is changing below 100K. In the temperature window where the bulk SR is expected the signal for both film thicknesses become more pronounced with a clear increase in the exchange field below 10K, the temperature regime where Dy ordering sets in. The fluctuations are probably the result of a competition between the $Dy^{3+}$ - $Fe^{3+}$ interaction leading to the SR and Fe spins partially compensating the FM Dy component while going through a SR. To yield a more pronounced exchange bias at higher temperatures the magnetic anisotropy needs to be enhanced to clamp the magnetization sufficiently. Consequently, the $LuFeO_3$ films with the non-magnetic Lu do not display any exchange bias, the measured moments are clearly smaller than for $DyFeO_3$ films, and $H_{C2}$ is almost a factor 10 smaller due to the absence of a $Lu^{3+}$-$Lu^{3+}$ interaction.

### 3.2.5 Magnetic exchange fields

To evaluate the ordering of the spin systems, the saturation moment for the 13nm film at 50 K for both spin system (hard and soft) is measured, which are ~1.445 $\mu_B$ and ~9.875 $\mu_B$ respectively and the maximum moments $Fe^{3+}$ and $Dy^{3+}$ can reach are 5.91 $\mu_B$ and 10.6 $\mu_B$, respectively (Fig 5d). Therefore, the soft magnetic ordering in $DyFeO_3$ films originates from the ordering of the Dy lattice. The soft magnetic ordering is also orientation and thickness dependent (Fig. 5a, b). Measuring $M_{sat}^{soft}(T)$ for [100] and [001] for the 115nm and 13nm film (Fig. 5c), it is clear that for the 13nm film [100] is the dominant direction, whereas for the 115nm film both directions are almost equal as one would expect it to be for a single crystal. A natural explanation is the large difference in growth induced strain in the 13nm film giving rise to the anisotropy for the moments as well as large differences in the absolute values for $\mu_B$. The smaller values for the thick film are closer to values measured for single crystals [7]. In Fig. 4a, b the dashed line at 50K indicates where a maximum or a plateau for the soft component appears. Comparing these measurements to the $M(T)$ measurements shown in Fig. 3, we note that transition temperatures identified at ~70K are related to the Fe spin reorientation, whereas the signature at ~50K is directly linked to Dy and hence the $Dy^{3+}$-$Fe^{3+}$ interaction. It is worth to mention that the constant presence of the DS loops at all measurable temperatures indicates the influence of Dy at room temperature and above. Besides, the large $M_{sat}^{soft}$ component in thinner films at room temperature shows that in highly strained films the influence of the $Dy^{3+}$ spins are more dominant as compared to thicker films.

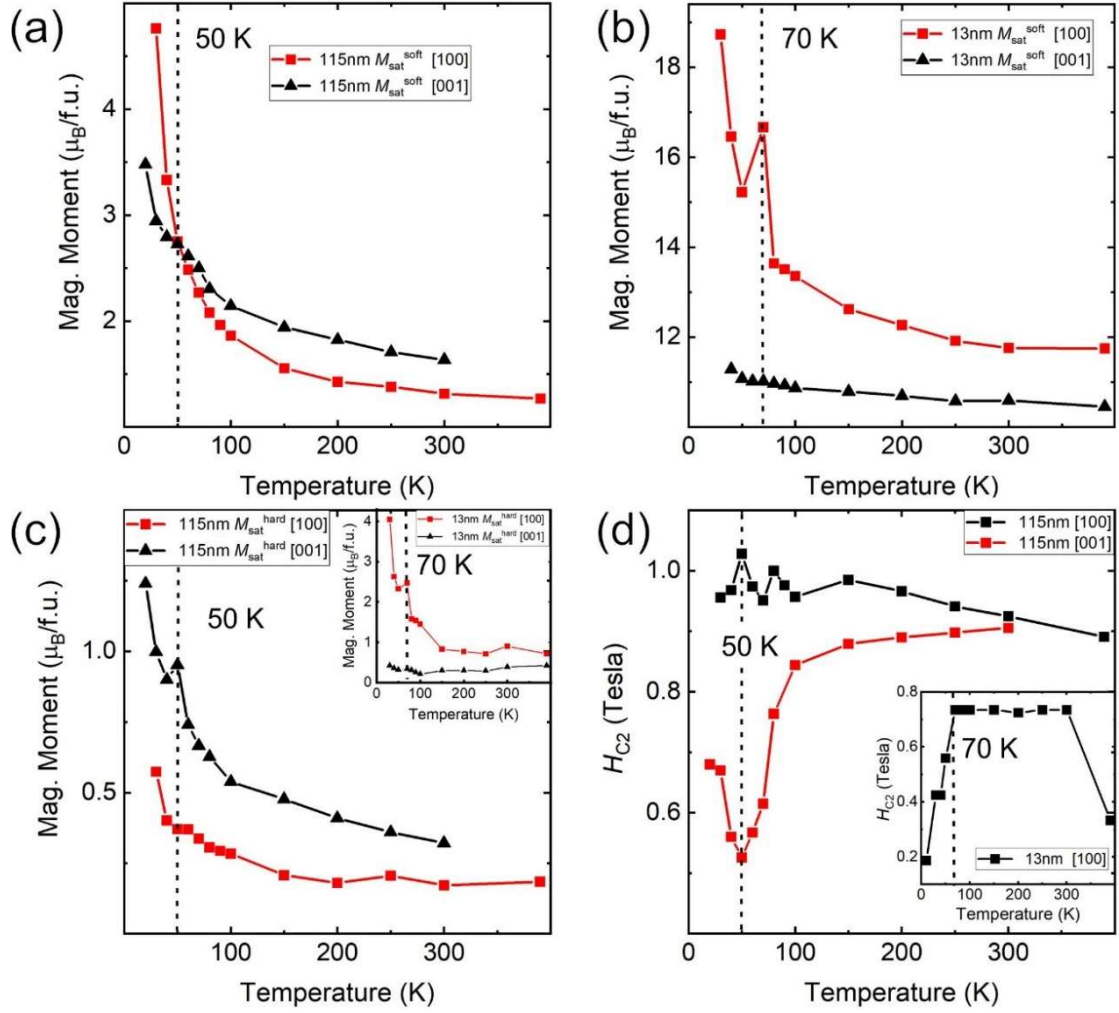

*Figure 5:* a) $M_{sat}^{soft}$ vs. $T$ of a 115nm DyFeO$_3$ film along [100] and [001]. b) $M_{sat}^{soft}$ vs. $T$ of a 13nm DyFeO$_3$ film along [100] and [001]. c) $M_{sat}^{hard}$ vs. $T$ of a 115nm film along [100] and [001]. Insert: $M_{sat}^{hard}$ vs. $T$ of a 13nm film along [100] and [001]. d) $H_{C2}$ vs. $T$ of a 115 nm film along [100] and [001]. Insert: $H_{C2}$ vs. $T$ of a 13nm film along [100];

So far, we have shown that the magnetic properties of strained DyFeO$_3$ thin films are orientation dependent and the Dy$^{3+}$-Fe$^{3+}$ interaction is noticeable up to 390K. Unlike in a DyFeO$_3$ single crystal, Dy$^{3+}$ is magnetically active along [100] and [001] in the entire temperature window above and below $T_{N,Dy}$. This is evident when measuring $H_{C2}(T)$ for the 13nm and 115 nm thin film (Fig. 5d). The coercive field $H_{C2}$ for the [100] direction is almost temperature independent and just below 1 T for the 115 nm film. For the [001] direction, it is smaller and has a distinct minimum at around $T_{SR}$ indicating that the Fe lattice undergoes a partial SR. However, the value of $H_{C2}$ partially recovers with further decreasing temperature. The coercive field $H_{C2}$ is also considerably smaller than the critical field needed to induce FE in bulk. The situation is similar for the 13nm film. Between 300 K and 70 K $H_{C2}(T)$ is temperature independent along [100] and starts to drop when SR occurs. Below 70K it is not possible to determine $H_{C2}$. For the [001] direction, no data can be extracted from $M(H)$ loops as the measured hysteresis cannot be clearly distinguished from a DS and a potential value for $H_{C2}$ would be similar to the maximum applied field of 5T used to measure $M(H)$. However, for the other film thicknesses investigated, the tendency was as shown exemplary for the thinnest and thickest film. The meaning of the identified extrema at 60K for [001] and 50K and 80K for [100] is not clear at present. These measurements show that [100] is the magnetically distinct and dominant direction with respect to the $Dy^{3+}$-$Fe^{3+}$ interaction as compared to [001] even well above room temperature. For single crystals no DS loop are measured for any crystalline direction in the temperature regime between room temperature and 10 K [7]. A possible explanation for the stronger influence of the $Dy^{3+}$ along [100] can be the result of the orbital overlap of the $Dy^{3+}$ with the oxygen in the unit cell due to the mixture of compressive and tensile strain.

## 4 Discussion

From the presented data on magnetic properties of thin orthoferrite films and single crystal, it is evident that the thin film properties are highly affected by the epitaxial lattice strain and hence very different compared to single crystal data. Most noticeable is the significant shift in $T_{SR}$, the very broad SR, or its complete suppression during FC measurements. The latter observation is an indication, that $\Gamma_4$ is the preferred magnetic space group also supported by a FM hysteresis measured between 1.6 K and 390 K.

When comparing the changes in film lattice parameter for all three directions with the pressure induced changes for a single crystal, the observed maximum changes for the lattice parameters would correspond to 12kbar for [100], 15kbar for [010] and 16.7kbar for the [001] direction. This is clearly more than expected for the two short axes, and considerably less for the long axis if applying the rule of a thumb criteria of 1% lattice mismatch corresponding to 10kbar (1GPa) epitaxial strain. Overall, these nominal pressure values are closer to a more uniform pressure distribution in the unit cell justifying the comparative approach studying magnetic properties of a single crystal under hydrostatic, uniform He-pressure.

How a small perturbation in the lattice parameters of the $DyFeO_3$ unit cell can lead to noticeable changes in its magnetic properties has already reported for single crystal and powder samples [7]. The increase in $T_{SR}$ with the amount of compressive epitaxial strain roughly corresponds to the compressive strain utilized in uniaxial pressure experiments on $DyFeO_3$ single crystals where a change in $T_{SR}$ of 2K/kbar was measured. Uniaxial measurements have the benefit when measuring materials properties in one specific direction, their pressure dependent changes will be reflected correctly. In films we typically have anisotropic strain and hence measure a mixture representative of the properties for the different crystallographic directions. Comparing the observed values of $T_{SR}$ with the changes in lattice parameters, we would expect to observe a minimum change on $T_{SR}$ of +20K and up to 40K based on the nominal pressure values. The observed window for $T_{SR}$ for the different films fits well to the expected temperature interval. Whereas the Morin transition along [001] remains sharp with increasing pressure for a single crystal SR, for films along [001] it is relatively broad (Fig. 2). This leads to the conclusion that the broadening of the SRT is the result of the anisotropic strain in these films.

Another obvious difference between single crystal and thin film properties is the presence of a FM hysteresis loop in *M*(*H*) measurements between 1.6K and 390K along [001]. For a single crystal, the FM hysteresis closes at $T_{SR}$ signalling the change from $\Gamma_4$ to $\Gamma_1$. This does not happen in films and confirms that a $\Gamma_4$ to $\Gamma_1$ transition does not occur in our films. In addition, a double hysteresis loop originated by the coupling between the Fe and Dy spins has been observed along [100] and [001], the probable origin of the DS in LuFeO$_3$ thin films as a single spin system is biaxial magnetic anisotropic strain in a unit cell [39-42], giving rise to a stronger Fe-O octahedra tilting and hence a canting along both in-plane directions. The DS is clearly more pronounced in the thicker film where strain along [100] changes from compressive to tensile biaxial anisotropy alike.

The presence of the DS in DyFeO$_3$ also reveals a large coercive field $H_{C2}$ of up to 1T. It is, however, smaller than the critical field of 2.3 T needed to induce FE in bulk. The combination of growth induced "pressure" and $H_{C2}$ is sufficient to drive a polar state in DyFeO$_3$ [24]. It is at present not clear if the polar state can be maintained sufficiently well in order to have ferroelectricity reliably or if depolarization effects come into play. The orientation dependence of $H_{C2}$ is also suggestive of [100] probably being the polar axis instead of [001] as reported for single crystals.

## 5 CONCLUSIONS

We have studied structural and magnetic properties of compressively strained orthorhombic (010)-oriented DyFeO$_3$ and LuFeO$_3$ thin films grown on (010)-oriented YAlO$_3$ and compared the results to the pressure dependence of the (031) magnetic Bragg peak for a DyFeO$_3$ single crystal under uniaxial strain. (010)-DyFeO$_3$ thin films grow coherent up to 40nm, with some lattice relaxation and a smaller unit cell volume than unstrained DyFeO$_3$. Thicker films relax towards bulk lattice values with the net unit cell volume larger than bulk. For those films strain is mostly tensile for all directions, for films thinner than 40nm compressive along [100] and [001] and tensile along [010]. For LuFeO$_3$, strain states are similar to DyFeO$_3$ for very thin films. For both materials, this is indicative of a Poisson ratio < 0.5 when grown as a thin film.

Shifting $T_{SR}$ for DyFeO$_3$ films by more than 20 K to higher temperatures or even completely suppressing the spin reorientation for thinner films (<40nm) with an applied magnetic field shows the presence of the $\Gamma_4$ phase. This is substantiated by measuring double-step hysteresis loops in the whole measured temperature range above $T_{N,Dy}$ and the presence of an induced magnetically ordered Dy lattice. As Dy orders antiferromagnetically, the DS hysteresis disappear and an exchange bias effect has been observed instead.

The magnetic field at which the DS occurs is of the order of 1T and could be large enough to drive the unit cell polar. Dy$^{3+}$ is magnetically active along [100] and [001] with the preference on the [100] direction. Since an ordering of the Fe lattice induces ordering in the Dy lattice and ferroelectricity in bulk is linked with a ferromagnetically ordered Dy lattice in the $\Gamma_4$ phase, it is expected to observe ferroelectricity along the [100] direction in DyFeO$_3$ thin films between $T_{N,Dy}$ and 390 K. For LuFeO$_3$ we also observe DS hysteresis loops along [100] due to a strain induced magnetic anisotropy. It is therefore likely to measure a polar response at room temperature and above along the *a*-direction as predicted by theory. Polarization measurements are required to show if magnetic anisotropy for both materials is the mechanism to induce FE, which would then be a generic mechanism for this class of materials.

**ACKNOWLEDGMENTS**

This work was supported by the Swiss National Science Foundation (Project No. 200020_169393) and the Paul Scherrer Institute. We also would like to acknowledge the help of V. Ukleev and D. Singh with single crystal neutron measurements on the Morpheus beam line.